\documentclass[epj-spec]{svjour}
\usepackage{graphics}
\begin{document}
\title{Control of Optical Dynamic Memory Capacity of an Atomic Bose-Einstein Condensate}
%\subtitle{Do you have a subtitle?\\ If so, write it here}
\author{Devrim Tarhan\inst{1}, Alphan Sennaroglu\inst{2}, \and \"Ozg\"ur E. M\"ustecapl{\i}o\u{g}lu\inst{2}}
\institute{Harran University, Department of Physics, 63300
Osmanbey Yerle\c{s}kesi, \c{S}anl\i{}urfa, Turkey \and Ko\c{c}
University, Department of Physics, Rumelifeneri Yolu, 34450
Sar{\i}yer, Istanbul, Turkey}
\abstract{Light storage in an atomic Bose-Einstein condensate is
one of the most practical usage of these coherent atom-optical
systems. In order to make them even more practical, it is
necessary to enhance our ability to inject multiple pulses into
the condensate. In this paper, we report that dispersion of pulses
injected into the condensate can be compensated by optical
nonlinearity. In addition, we will present a brief review of our
earlier results in which enhancement of light storage capacity is
accomplished by utilizing multi-mode light propagation or choosing
an optimal set of experimental parameters.}
%end of abstract
%
\maketitle

\section{Introduction}

Soon after the generation of Bose Einstein condensate in an
ultracold gas of trapped Alkali atoms \cite{bec95}, a promising
utilization of it for the dramatic slowing down of a light pulse
was demonstrated \cite{slowlight-exp}. This feat is accomplished
with the help of a quantum coherent effect called
electromagnetically induced transparency (EIT) \cite{eit1,eit2}.
Ultraslow light pulses can be used for storage of coherent optical
information \cite{lightstorage}. Mutual conversion of classical
coherent information (phase and amplitude) of the light pulse and
the quantum information (quantum state) of the atomic system can
allow for quantum information processing via ultraslow light
\cite{dutton}. A quite recent experiment provides strong hope
towards this direction \cite{ginsberg}. In the experiment,
coherent optical information is first encoded in one condensate.
Quantum information in the condensate is then carried to another
condensate by a matter wave. Finally, coherent optical information
is revived in an optical pulse, generated out of the new
condensate on demand.

Due to such impressive developments in experimental ability to
control of light and matter waves in light storage experiments,
more practical quantum information processing applications can be
expected to occur in near future. On the other hand, it is
necessary first of all, to increase our capability to control the
amount of information stored in the condensate. For that aim, we
should be able to inject more than one pulse into the condensate
during the storage time. This is a basic requirement to realize
practical logic gates in atomic condensates using slow light set
ups \cite{lightstorage,dutton}.

To investigate how to make more than one pulse simultaneously
present in an atomic condensate efficiently, we have performed a
series of studies. This paper reviews some of our earlier results
and in addition it reports our new results where we have found
optimum conditions for dispersion compensation via nonlinearity.
In our earlier works, we show that optical dynamic memory capacity
of the condensate can be optimized by choosing a certain set of
experimental control parameters, such as coupling laser Rabi
frequency and temporal width of the probe pulse for a given atomic
condensate \cite{tarhan}. It has been shown that axial density
profile of the condensate helps to preserve the probe pulse shape
against group velocity dispersion \cite{tarhan}. Further
enhancement of the memory capacity can be possible by taking into
account radial confinement of the probe pulse. We have
demonstrated that a particular radial density profile of the
condensate reduces the effect of modal dispersion and contributes
to the pulse shape preservation \cite{tarhan2}. We deduce that the
optical control parameters are beneficial for optimizing the
memory capacity and at the same time, properties of the atomic
cloud can also be exploited to enhance the maximum capacity
available. When we take into account radial confinement, ultraslow
wave-guiding regimes have been investigated and characterized.
Propagation constants and the conditions on the number of
ultraslow optical modes have been determined. Detailed numerical
examinations revealed the ultraslow mode profiles that can be
supported by atomic condensates. In this paper, we will go beyond
these linear optical results and examine the effects of nonlinear
optical response of the condensate. We show that ultraslow
solitons can bring further enhancement in the light storage
capacity.

Following this introduction, in Sec. 1, the paper first reviews
optical information storage enhancement schemes based upon linear
optics considerations. In Sec. 2.1., group velocity dispersion and
a scheme to beat it are discussed. In Sec. 2.2, taking into
account transverse directions to the propagation direction, modal
dispersion is studied. Information storage enhancement with the
aid of ultraslow waveguiding modes are discussed. Our new
contributions in which ultraslow light propagation is studied in
the nonlinear regime is described in Sec. 3. Increase in the
capacity of optical information storage due to the ultraslow
optical solitons is discussed in that section. Finally, we
conclude in Sec. 4.

%%%%%%%%%%%%%%%%%%%%%%%%%%%%%%%%%%%%%%%%%%%%%%%%%%%%%%%%%%%%%%%%%%%%%
%%%%%%%%%%%%%%%%%%%%%%%%%%%%%%%%%%%%%%%%%%%%%%%%%%%%%%%%%%%%%%%%%%%%%

\section{Control of optical information storage in linear optical regime}

To slow down group velocity of an optical pulse just to few meter
per second, steep dispersion of electromagnetically induced
transparency is used \cite{slowlight-exp}. Optical pulse to be
slowed down consists of a weak probe light. It is shined upon
atomic BEC, together with a stronger control field. This
configuration allows for cancellation of absorption at resonance
and creates transparency over a small frequency window. Measuring
the axial length of the condensed cloud by an imaging laser and
comparing the time delay in the probe beam propagation with a
reference beam not entered into the condensate determines
operationally the average group speed of the probe pulse. Such a
typical slow light experiment is usually modelled by a one
dimensional linear optics model. This model predicts the measured
group speeds well, especially when the interactions within the
cloud is taken into account \cite{ozgur}.

Considering a gas of effectively three-level atoms, linear optical
response of an EIT medium is characterized by a susceptibility
given by
\begin {equation}
\label{chieit} \chi = \frac{\rho|\mu_{31}|^2}{\epsilon_0 \hbar}
\frac{i(i \Delta + \Gamma_2/2)}{[(\Gamma_2/2 + i
\Delta)(\Gamma_3/2 + i \Delta) + \Omega_c^2/4]}.
\end {equation}
The effective level configuration of the atoms consists of two
dipole allowed transitions such that the lower levels $|1\rangle$
and $|2\rangle$ are coupled to the upper level $|3\rangle$ by the
probe and the control fields, respectively. Here, $\rho$ is atomic
(number) density. We assume that the control field of Rabi
frequency $\Omega_c$ is at resonance, while $\Delta$ is the
detuning of the probe field of frequency $\omega$ from the
resonance frequency $\omega_0$. $\mu_{31}$ is the dipole matrix
element for the probe transition. Taking resonance wavelength
$\lambda$ for the probe transition we have the usual relation
$\mu_{31}^2=3\epsilon_0\hbar\lambda^3\gamma/8\pi^2$ with $\gamma$
is the spontaneous decay rate for the probe transition. Dephasing
rates of the corresponding transitions are denoted by $\Gamma_2$
and $\Gamma_3$. Within a small frequency window about the probe
resonance, susceptibility becomes a negligibly small pure
imaginary number, which leads to the transparency effect. Besides,
within this window, susceptibility shows steep variation with
$\Delta$ so that group velocity reduces dramatically for a
resonant probe pulse that can propagate without suffering any
significant absorption. Physically, this is possible by the
cancellation of absorption due to the destructive quantum
interference of the transition probabilities to the upper level.
Quantum coherence is established by the control field which is
stronger than the probe field in the usual EIT condition.

It may be argued that BEC is not essential for the ultraslow light
generation. Degenerate Fermi gas also allows for significant
reduction in group velocity, though not as efficient as BEC
\cite{ozgur2}. BEC gives much slower speeds over more compact
spatial dimensions. Its unique advantage however is the ability to
combine matter wave dynamics with the ultraslow light pulses
\cite{ginsberg}. This seems to be more promising for practical
quantum information processing.

Susceptibility formula is derived by the steady state solution of
the Liouville equation for the density matrix, assuming most of
the atoms remain in level $|1\rangle$. In the subsequent section,
we will go beyond linear regime and consider nonlinear response.
In the linear regime, we will first discuss dispersive propagation
of ultraslow short pulses in one dimension. After that, we extend
this by taking into account transverse directions as well.

Dispersive effects are weak in the usual ultraslow light
experiments as these pulses are just a few microseconds long.
First attempts to model experimental system uses one dimensional
non-interacting BEC as the propagating medium with linear optical
response \cite{agarwal}. Furthermore it is assumed that the medium
is non-dispersive. This model is improved significantly by
considering atom-atom interactions. It has been shown that
increasing interactions lead to faster group speeds \cite{ozgur}.
Higher dimensional propagation effects within paraxial
approximation are also considered and diffraction effects at the
thermal cloud boundary of the condensate are numerically simulated
\cite{ozgur}. These results demonstrate validity of the
non-dispersive, one-dimensional model in describing experimental
results. On the other hand, when it comes to improve storage
capacity, this model becomes insufficient.

Natural attempt to enhance our ability to inject multiple pulses
into the BEC during the storage time is to use shorter pulses in
time. By this way, if we say every pulse carries a bit of coherent
optical information (which is the classical information associated
with the amplitude and phase of the pulse), then we could achive
higher bit storage capacity. With the usual microsecond pulses
however, as the storage time is also about a few microseconds, one
can get only one pulse present in the condensate
\cite{slowlight-exp}. Some schemes are proposed to increase number
of pulses. One can utilize for instance different polarization of
probe pulses \cite{agarwal02} or enhancement of EIT window
\cite{wei05}. Both these proposals would benefit from more direct
approach of being able to inject shorter pulses into the
condensate \cite{sautenkov05,sun05,bashkansky05}. The fundamental
problem in using shorter pulses is that they may be broadened due
to the group velocity dispersion within the storage time. It is
thus necessary to study effect of group velocity dispersion in the
axial propagation of ultraslow short pulses in order to estimate
actual enhancement of the bit storage capacity.

\subsection{Group velocity dispersion}

We will now investigate the propagation of short laser pulses
under slowly varying phase and envelope approximations without
taking into account nonlinear optical response. A formulation of
light propagation through a dispersive EIT medium is given in Ref.
\cite{dispersive}. We have applied this theory to ultraslow short
light pulses through BEC in Ref. \cite{tarhan}. We now review and
summarize our major results in Ref. \cite{tarhan}.

Total polarization is related to the electric field through
electric susceptibility via $\mathbf{P}=\mathbf{P^{(l)}}$ where
the linear polarization is $\mathbf{P^{(l)}} = \epsilon_{0}
\chi(\omega)^{(1)}\mathbf{E}$. Dispersive medium is characterized
by a frequency dependent susceptibility. We write
$P(\omega-\omega_0,E)=\epsilon_{0}\chi(\omega-\omega_0)^{(1)}E(\omega-\omega_0)$.
Expanding the linear susceptibility $\chi^{(1)}$ of the dressed
atom to second order about a central frequency
$\omega_0$,\cite{dispersive}
\begin{eqnarray} \label{chi}
\chi(\omega-\omega_{0})^{(1)} &=& \chi(\omega_{0})
+\frac{\partial\chi}{\partial\omega}|_{\omega_{0}}(\omega-\omega_{0})
+\frac{1}{2}\frac{\partial^{2}\chi}{\partial^{2}\omega}|_{\omega_{0}}(\omega-\omega_{0})^2,
\end{eqnarray}
polarization becomes
\begin{eqnarray}\label{dispersifpolarization0}
P(t)  &=& \epsilon_{0}\chi(\omega_{0})^{(1)}E(t)
-i\epsilon_{0}\frac{\partial\chi^{(1)}}{\partial\omega}|_{\omega_{0}}
\frac{\partial E}{\partial t}-
\frac{\epsilon_{0}}{2}\frac{\partial^{2}\chi^{(1)}}{\partial^{2}\omega}|_{\omega_{0}}
\frac{\partial^2 E}{\partial^2 t}.
\end{eqnarray}
In slowly varying approximation, the wave-equation is given by,
\begin{eqnarray} \label{waveeq0}
\frac{\partial E}{\partial z} + \frac{1}{c} \frac{\partial
E}{\partial t} = \frac{i k}{2\varepsilon_0}P.
\end{eqnarray}
Using equation (\ref{dispersifpolarization0}) in Eq.
\ref{waveeq0}, we obtain the wave equation for the short optical
pulse such that
\begin{equation} \label{eq:nonlinearpulse}
\frac{\partial E}{\partial z} + \alpha E + \frac{1}{v_g}
\frac{\partial E}{\partial t} + i \,b_{2} \frac{\partial^2
E}{\partial t^2},
\end{equation}
where $\alpha$ determines the pulse attenuation; $v_g$ is the
group velocity, $b_{2}$ is the second-order dispersion
coefficient, or the group velocity dispersion. Calculating the
third and the second order dispersion coefficients using typical
experimental parameters \cite{slowlight-exp} but for shorter
pulses, we have found that third order dispersion is seven orders
of magnitude smaller and negligible. Second order dispersion
becomes significant for pulses shorter than microseconds. In our
numerical calculations, we consider parameters for the experiment
in Ref. \cite{slowlight-exp} except for the coupling field Rabi
frequency which is chosen to be $\Omega_c=5\gamma$. The
transmission window, $\Delta \omega$ is calculated to be $\Delta
\omega=8.49\times10^7$ Hz so that we shall consider pulses of
width up to $\tau=10$ ns for efficient transmission. The
coefficients can be calculated directly from the susceptibility
via
\begin{eqnarray}
\alpha &=& -\frac{i \pi}{\lambda} \chi(\omega_{0}),\\
\frac{1}{v_g} &=& \frac{1}{c} - \frac{\pi}{\lambda} \frac{\partial
\chi}{\partial \omega}|_{\omega_{0}},\\
b_{2} &=& \frac{\pi}{2\lambda}
\frac{\partial^{2}\chi}{\partial\omega^2}|_{\omega_{0}}.
\end{eqnarray}
These expressions lead to complex parameters in general, as $\chi$
is a complex valued function. At EIT resonance, they lead to
physically meaningful well-defined real valued absorption
coefficient, group velocity and dispersion coefficient. When Rabi
frequency for the coupling field is sufficiently large such that
$\Omega_c\gg \Gamma_{2,3}$, the coefficients reduce to
\begin{eqnarray}
\alpha &=& \frac{2\pi\rho|\mu_{31}|^2\Gamma_2}
{\epsilon_0\hbar\lambda\Omega_c^2},\\
v_g &=&
\frac{c\epsilon_0\hbar\Omega_c^2}{2\omega_{31}|\mu_{31}|^2\rho},\label{vg}\\
b_2&=&i\frac{8\pi\Gamma_3|\mu_{31}|^2\rho}{\epsilon_0\hbar\lambda\Omega_c^4}.\label{b2}\\
\end{eqnarray}
It should be noted that the given expression for for $v_g$ is
valid only when
\begin{eqnarray}
\Gamma_{2,3}\ll\Omega_c\ll\sqrt{4\pi
c|\mu_{31}|^2\rho/\epsilon_0\hbar}
=\sqrt{\frac{3}{4\pi}c\lambda^3\gamma\rho}.
\end{eqnarray}
From the explicit expressions, we see that $v_g$ increases more
slowly with the Rabi frequency of the control field $\Omega_c$
than the decrease of the dispersion coefficient $b_2$. This higher
susceptibility of $b_2$ to $\Omega_c$ can be exploited to optimize
the bit storage capacity $C$, defined by $C=L/2v_g\tau$ with
$\tau$ is the largest width of the probe pulse within the
condensate of size $L$.

Analytical solutions for $C$ are derived for a Gaussian pulse
assuming a uniform effective dispersive medium representing the
actual spatially inhomogeneous BEC \cite{tarhan}. Analytical
solution of the wave equation allows for determination of the
pulse width. $L$ is calculated as the root-mean-square of the
axial distance for a given condensate of number density $\rho$,
under a semi-ideal model of the weakly interacting condensate
\cite{naraschewski}. This leads to
\begin{eqnarray}
C=\frac{L}{2\tau_0\sqrt{4\pi^2\Omega_c^4/9\lambda^4\gamma^2\rho^2
+4L^2\Gamma_3^2/\pi^2\tau_0^4\Omega_c^4}}.
\end{eqnarray}
Initial pulse width is denoted by $\tau_0$. $C$ shows a maximum at
a critical Rabi frequency determined by
\begin{eqnarray}
\Omega_{c0}=\left(\frac{3\Gamma_3\lambda^2\gamma\rho
L}{\pi^2\tau_0^2}\right)^{1/4}.
\end{eqnarray}
The corresponding storage time of the pulse $t_{s0}$ is given by
\begin{eqnarray}
t_{s0}=\tau_0\left(\frac{\sqrt{3}\lambda\sqrt{\gamma}}{2\sqrt{\Gamma_3}}\right)
\left(\sqrt{\rho L}\right).
\end{eqnarray}
From the number of pulses that can be present in this storage time
we find the maximum capacity as,
\begin{eqnarray}
C_{max}=\sqrt{\frac{3\gamma\lambda^2L\rho}{32\Gamma_3}},
\end{eqnarray}
which is independent of the pulse width at the critical Rabi
frequency. At the critical Rabi frequency, about two orders of
magnitude improvement is achieved regarding the bit storage
capacity.

\begin{figure}[h]
\resizebox{0.50\columnwidth}{!}{
\includegraphics{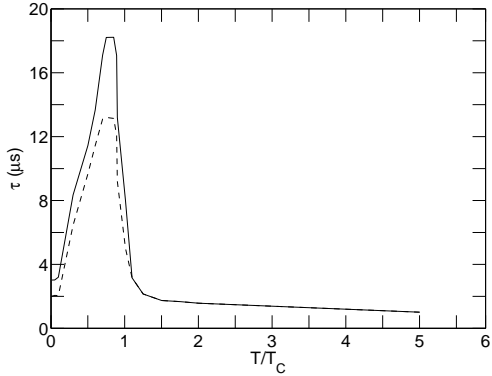}}
\caption{Temperature dependence of the pulse width calculated
numerically for an inhomogeneous condensate profile. Initial width
of the pulse is $1$ ns. Effect of the interactions can be seen by
comparing the solid curve for which the s-wave scattering length
characterizing the effective atom-atom interactions $a_s=2.75$ nm
with the dashed one for which $a_s=7$ nm. The parameters used in
the calculation are taken from the experiment in Ref.
\cite{slowlight-exp}. \label{fig1}}
\end{figure}

More rigorous numerical calculations are verified these analytical
results. Numerical calculations take into account density
variations and the thermal component of the gaseous system.
Temperature and the interactions affect the pulse character
through the dependence of pulse parameters on spatial size and
density product. Numerical studies are required to handle
spatially inhomogeneous BEC. Propagation parameters in the wave
equation gain non-local character under this model but can be
treated numerically. We have found that as the pulse gets broader
and broader as it propagates towards to the central region of the
atomic cloud, where the dispersive effects are most strong, it
becomes better protected against dispersion. As it is already
broadened, it gets less broadening effect. Axial density profile
of the condensate protects the pulse against group velocity
dispersion.  Thermal variation of the bit storage capacity is
investigated numerically and just below critical temperature, an
optimum temperature for the highest capacity is determined.
Interactions in the cloud effects the broadening of the pulse and
makes it more sharp when the interactions are stronger. Fig.
\ref{fig1} illustrates these results. %%

From the figure we see that the short pulses suffer significant
broadening despite the short size of the condensate. One cannot
increase the capacity $C$ thousand times by using nanosecond
pulses instead of microsecond ones. Using a control field at the
suggested critical value as above, an optimum increase of the
number of pulses that can be injected multiply into the condensate
can be achieved. Analytical result of $C$ evaluated gives about
$80$ pulses in the condensate while numerical result suggests, as
explained above, due to less broadening within the inhomogeneous
condensate, $800$ pulses could in fact be stored. In comparison to
present slow light experiments, that means a two orders of
magnitude enhancement in the bit storage capacity. Below we shall
discuss additional improvements to $C$, taking into account
transverse directions of the condensate as well as nonlinearity in
the probe pulse propagation.

\subsection{Modal dispersion}

In a subsequent analysis we take into account transverse
dimensions to the axial propagation direction and examine the
ultraslow optical modes. Waveguiding problems in cold gaseous
systems was subject to much attention recently
\cite{duan,andre,cheng,veng}. Now we summarize our major results
reported in Ref. \cite{tarhan2}. We observe that at extremely low
temperatures well below the critical temperature of condensation,
density profile of the condensate can be approximately described
by Thomas-Fermi approximation. Normally, at perfect EIT resonance,
no significant index difference between the condensate core and
the thermal background gas can be achieved. In this case, only
single mode is supported in the condensate. For a slightly
off-resonant probe however, sufficiently high index contrast can
be generated between the condensed core and the enveloping thermal
gas. Under these conditions, radial density of the condensate can
be translated to a graded index profile characterized by a
parabolic (quadratic) spatial variation in the form
\begin{eqnarray}
n(r) = \cases{n_1[1-A(\frac{r}{R})^2]^{1/2} & $r\leq R$. \cr 1 &
$r\geq R$},
\end{eqnarray}
with $A=1-1/n_1^2$. Here $R$ is the thermal radius along the
transverse (radial, $r$) direction and $n_1$ is the index along
the axial ($z$) direction. Such a system resembles a graded index
fiber, though one of micron sized. Light propagation is found to
be in the weakly guided regime where the optical modes are
described in terms of linearly polarized (LP) modes as the index
difference is small and the axial component of the electric field
is negligible. In this case the radial wave equation is described
by the Helmholtz radial equation
\begin{eqnarray}
\label{diffequa} \left[\frac{{\rm d}^2}{{\rm
d}r^2}+\frac{1}{r}\frac{{\rm d}}{{\rm d}r} +k_{0}^2 n^2(r)-\beta
^2-l^2/r^2\right ]\psi(r)=0.
\end{eqnarray}
Here $k_0=\omega/c,l=0,1,2,...$, $\beta$, and $\psi$ are defined
for the transverse field of the LP modes $E_t=\psi(r)\exp{[{\rm
i}(l\phi+\omega t-\beta z))]}$.

\begin{figure}[h]
\resizebox{0.50\columnwidth}{!}{
\includegraphics{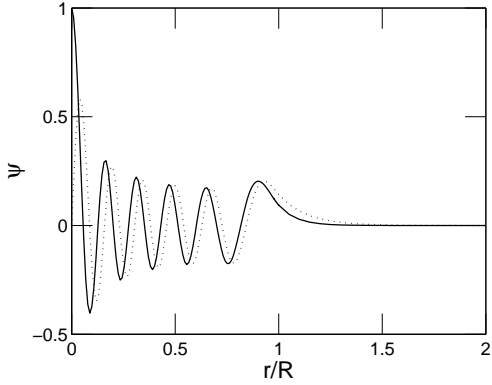}}
\caption{Two ultraslow transverse modes confined in a BEC. Solid
curve is for LP$_{00}$ while the dotted one is for LP$_{10}$. Mode
profiles are calculated with the parameters that are taken from
the experiment in Ref. \cite{slowlight-exp}, except for
$\Delta=-0.1\gamma$ and $\Omega_c=2.5\gamma$. \label{fig2}}
\end{figure}

The number of such transverse (LP) modes as well as their group
velocity is evaluated. We find that modal dispersion is stronger
at low temperatures and large number of modes can be supported by
the condensate. As the temperature gets higher and higher, number
of modes that can be supported decrease. This is due to the
shrinkage of the condensate and the vanishing of the index
contrast. Analytical results are provided by a conventional WKB
treatment. Conditions for multiple, single and two mode
propagations are provided. It is shown that all LP modes can
propagate at ultraslow light speeds. Lower order modes propagate
slower while higher order modes goes faster as they are away from
the central, more dense, condensed region. This particular density
profile of the BEC then allows for less pulse spread due to
differences in the group speeds of different modes. As in the case
of axial propagation where axial density of the condensate
protects the pulse against group velocity dispersion, now the
radial profile of the condensate also helps for the pulse shape
preservation against modal  (waveguide) dispersion. More detailed
numerical studies are performed. We consider the condensate in
terms of tiny shells and treat the index as constant within each
region between the shells. At constant index, analytical solutions
in terms of Bessel function are written down and matched at the
boundaries. By this way, actual mode profiles are numerically
determined. We illustrate two modes in Fig. \ref{fig2}. %%

Short axial size of BEC and its unique density profile can be
exploited to keep modal dispersion weak. In such short distance
applications, this may be utilized to capture optimum amount of
power from a light source and enhance the optical coupling/sensing
capability of BEC.

LP modes can be addressed by specifically constructed transverse
profiles of the probe pulses. This can be exploited to control of
capacity in the transverse direction for storage of coherent
optical information in different mode patterns. These results can
be further combined with the longitudinal control of optical
information storage capacity where a particular EIT scheme, in
which the control field is perpendicular to the probe field, is
employed \cite{dutton2}.

\section{Control of optical information storage in nonlinear optical regime}

We will now take into account nonlinear optical response of the
condensate in the short pulse propagation. Total polarization, now
including the contribution of nonlinear polarization
$\mathbf{P^{(nl)}}= \epsilon_{0}(
\chi(\omega)^{(2)}:\mathbf{E}\mathbf{E}+\chi(\omega)^{(3)}:\mathbf{E}\mathbf{E}\mathbf{E})$,
can be expressed as
\begin{eqnarray} \label{polarization}
P(\omega-\omega_0,E)&=&\epsilon_{0}\chi(\omega-\omega_0)^{(1)}E(\omega-\omega_0)
+\epsilon_{0}\chi^{(2)}:\mathbf{E}\mathbf{E}+\epsilon_{0}\chi^{(3)}:\mathbf{E}\mathbf{E}\mathbf{E},
\end{eqnarray}
where an expression for the third order nonlinear susceptibility
is given in Ref. \cite{prokhorov}. This introduces nonlinear terms
to the susceptibility expansion so that polarization can be
written as
\begin{eqnarray}\label{dispersifpolarization1}
P(t)  &=& \epsilon_{0}\chi(\omega_{0})^{(1)}E(t)
-i\epsilon_{0}\frac{\partial\chi^{(1)}}{\partial\omega}|_{\omega_{0}}
\frac{\partial E}{\partial t} \nonumber-
\frac{\epsilon_{0}}{2}\frac{\partial^{2}\chi^{(1)}}{\partial^{2}\omega}|_{\omega_{0}}
\frac{\partial^2 E}{\partial^2
t}\nonumber\\&+&\epsilon_{0}\chi^{(2)}:\mathbf{E}\mathbf{E}+\epsilon_{0}\chi^{(3)}:\mathbf{E}\mathbf{E}\mathbf{E}.
\end{eqnarray}
$\chi^{(2)}$ is zero due to the inversion symmetry of BEC. Using
Eq. (\ref{dispersifpolarization1}) in Eq. (\ref{waveeq0}), we get
a nonlinear wave equation for the short optical pulse such that
\begin{equation} \label{eq:nonlinearpulse}
\frac{\partial E}{\partial z} + \alpha E + \frac{1}{v_g}
\frac{\partial E}{\partial t} + i \,b_{2} \frac{\partial^2
E}{\partial t^2} + i \eta |E|^2 E   = 0,
\end{equation}
where  $\eta$ is the nonlinear optical Kerr coefficient that can
be calculated from the susceptibility via
\begin{eqnarray}
\eta&=-&\frac{\pi}{\lambda}\chi^{(3)}(\omega_{0}).
\end{eqnarray}
When $\Omega_c\gg \Gamma_{2,3}$, we find
\begin{eqnarray}
\eta=\frac{4\pi\rho |\mu_{31}|^4 \Gamma_2 }{3 \varepsilon_0
\hbar^3
\lambda\Gamma_3}\frac{(\Gamma_2+\Gamma_3)}{\Omega_c^4}\label{eta}
\end{eqnarray}
Let us now write the wave equation (\ref{eq:nonlinearpulse}) in a
more conventional form \cite{nonlinearfiber},
\begin{equation} \label{eq:nonlinearpulse1}
\frac{\partial E}{\partial z} + \frac{\alpha^{'}}{2} E + \beta_1
\frac{\partial E}{\partial t} + \frac{i}{2}\beta_{2}
\frac{\partial^2 E}{\partial t^2} +i \eta |E|^2 E   = 0.
\end{equation}
Here $\beta_1=1/v_g$, $\beta_2=2 b_2$ and $\alpha^{'}=2\alpha$. By
making transformation $T=t-z/v_g$, $A=\sqrt{\epsilon_0 c}E$ and
$\gamma=\eta/\epsilon_0 c$ \cite{nonlinearfiber} we obtain
\begin{equation} \label{eq:nonlinearpulse2}
i\frac{\partial A}{\partial z} + i\frac{\alpha^{'}}{2} A -
\frac{1}{2}\beta_{2} \frac{\partial^2 A}{\partial T^2} - \gamma
|A|^2 A   = 0.
\end{equation}
where $A$ is the slowly varying amplitude of the pulse envelope.

\begin{figure}[h]
\resizebox{0.50\columnwidth}{!}{
\includegraphics{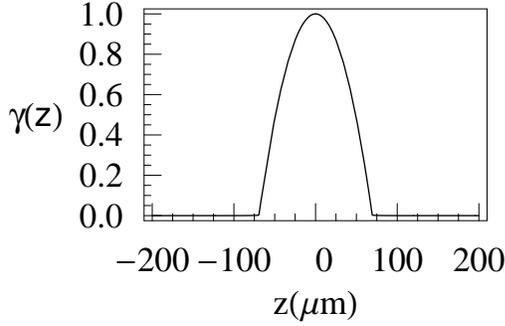}}
\caption{Axial spatial profile of the nonlinear coefficient
$\gamma$ for the parameters same with those of Fig.\ref{fig1}.
$\gamma$ is normalized by its peak value $3.64\times10^{-6}$ m/W.
\label{fig3}}
\end{figure}

\begin{figure}[h]
\resizebox{0.50\columnwidth}{!}{
\includegraphics{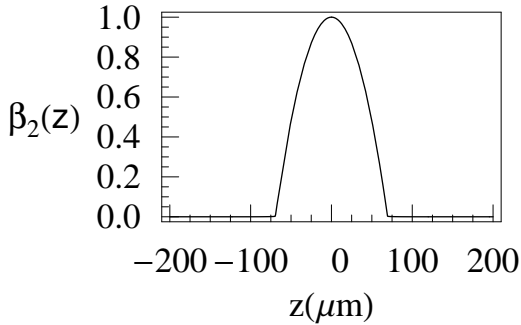}}
\caption{Axial spatial profile of the second-order dispersion
coefficient $\beta_2$ for the parameters same with those of
Fig.\ref{fig1}. $\beta_2$ is normalized by its peak value
$1.07\times10^{-11}$ s$^2$/m. \label{fig4}}
\end{figure}

Spatial variation of the nonlinear optical Kerr coefficient in the
axial direction is shown in Fig. (\ref{fig3}).  Fig. (\ref{fig4})
displays the axial spatial profile of the second-order dispersion
coefficient. %%
These figures demonstrate a potential for compensation of the
dispersion by nonlinearity is possible. We need to determine
required laser power for that aim. %%

In order to solve Eq. (\ref{eq:nonlinearpulse2}) one can introduce
normalized amplitude $U$ by using the definition
\cite{nonlinearfiber}
\begin{equation} \label{probefield}
A(z,\tau)=\sqrt{P_0}exp(-\alpha^{'}z/2)U(z,\tau).
\end{equation}
where $P_0$ is the peak intensity of the incident pulse and
$\tau=T/T_0$ with initial pulse width $T_0$. The dispersion length
$L_D$ and the nonlinear length $L_{NL}$ provide length scales over
which dispersive or nonlinear effects become important for the
pulse evolution in the Bose-Einstein condensate with an effective
length $L$. The dispersion length $L_D$ and the nonlinear length
$L_{NL}$ are given by \cite{nonlinearfiber}
\begin{eqnarray}
L_D &=& \frac{T_0^2}{|\beta_2|} ,\\
L_{NL} &=& \frac{1}{\gamma P_0} .
\end{eqnarray}
For the nonlinear pulse propagation, a dimensionless parameter $N$
is introduced as
\begin{equation} \label{parameter}
N^2=\frac{L_D}{L_{NL}}=\frac{\gamma P_0 T_0^2}{|\beta_2|}.
\end{equation}
For values $N\approx1$ both Self-phase modulation(SPM) and
Group-velocity dispersion(GVD) play an equal role
\cite{nonlinearfiber}. Under this condition, we determine the peak
power to compensate the group-velocity dispersion for a short
optical pulse propagating in the BEC using
\begin{equation} \label{peakpower}
P_0= \frac{|\beta_2|}{T_0^2 \gamma}.
\end{equation}
Using Eqs. (\ref{b2}) and (\ref{eta}), we find the relation
between the probe pulse characteristics and the material
parameters such that
\begin{eqnarray}
P_0T_0^2=\frac{32\pi^2\hbar
c}{\lambda^3}\frac{\Gamma_3}{\gamma\Gamma_2}.
\end{eqnarray}
Using the parameters given in the slow light experiment
Ref.\cite{slowlight-exp}, such that $\gamma/2\pi=10.01$ MHz,
$\lambda=589.76$ nm, $\Gamma_3=\gamma/2$, $\Gamma_2/2\pi=1$ kHz,
we calculate that $P_0=389$ mW/cm$^2$ for a $1\,\mu$s pulse. For a
shorter pulse of $10$ ns, required intensity increases to $3.9$
kW//cm$^2$. These intensities are beyond the saturation intensity
of the corresponding transition for the sodium atom, which is in
the order of $\sim 10$ mW/cm$^2$. However, it is common to
consider high intensities, in particular for the control field in
EIT. For example, Ref. \cite{prokhorov} considers a control field
of intensity $55$ mW/cm$^2$. In the EIT process, absorption of the
probe field is cancelled due to the quantum interference of the
transition amplitudes. The major limitation on the probe intensity
is the EIT condition itself, which requires much higher control
field Rabi frequencies than that of the probe field. If we use the
same parameters with Ref. \cite{prokhorov}, which takes slightly
different decoherence and decay rates than the ones used in the
experiment in Ref. \cite{slowlight-exp}, we would get $P_0=1.6$
mW/cm$^2$ for a $1\,\mu$s pulse and $P_0=166$ mW/cm$^2$ for a
$0.1\,\mu$s pulse. Sensitivity of the required peak intensity on
the natural linewidths and the decoherence rate of the atomic
system suggest that one can engineer linewidths using different
atoms, level structures \cite{wielandy} and putting the atoms in
an optical cavity \cite{cavityeit} to optimize demanded peak
intensity. Furthermore, instead of a continuous wave (cw) shining
upon the condensate, we consider storage of pulse trains inside
the condensate, and hence the average power would be smaller than
the one associated with the peak intensity. The results reported
here can be extended to the ultraslow pulses in hot gases as well,
by taking into account Doppler and collision broadenings. In such
systems, much higher pulse intensities in the orders of $1-1000$
W/cm$^2$ are used in EIT experiments \cite{hotgas}.

We would like to emphasize that other applications based upon
nonlinear effects might have different and less restrictive
requirements on the peak intensity or power of the probe pulse. As
a low loss medium with enhanced nonlinearity \cite{wang}, with
small spot size due to small condensate cross section,
self-focusing effects and EIT on the probe beam focusing, many
nonlinear effects of various interest could be realized with low
powers in the condensates in the EIT configuration. Our focus and
the purpose here is to examine the standard method of dispersion
compensation via nonlinearity so that number of pulses that can be
simultaneously present in the condensate can be optimized more
directly. In the earlier sections, we have discussed indirect
methods based upon adjusting control field intensity as well as
taking advantage of higher dimensional pulse propagation and the
transverse optical modes inside the condensate to enhance the
optical storage capacity. The present section demonstrates the
possibility of exploiting nonlinear compensation. With less
demanding probe pulse power requirements, rich physical situations
can be created such as quantum correlations in the weak probe
induced by the nonlinearity \cite{prokhorov}. From classical point
of view the relatively simpler theory of nonlinear pulse
propagation is employed here. Beyond dispersion compensation, such
a simple and well developed theory can also lead to rich
applications and some novel effects, in particular in the context
of dark and bright ultraslow soliton propagation
\cite{soliton1,soliton2,soliton3}, coupled matter wave and optical
field nonlinear dynamics, and so on. However their discussion is
beyond the scope of the present paper, which is limited to optical
storage capacity enhancement in BECs.

\section{Conclusion}

We have investigated the problem of injecting multiple ultraslow
pulses through a BEC. We have examined three methods for that aim.
This paper first reviews our two earlier results where nonlinear
optical response is ignored, then also presents our new results
where nonlinearity is present. In our first study, we have
considered short ultraslow pulse propagation in one dimension.
Taking into account group velocity dispersion, an optimum set of
experimental parameters, in particular Rabi frequency of control
field, that would give maximum pulse storage capacity are
determined. Secondly, transverse direction is taken into
consideration, and waveguiding of ultraslow pulses is studied.
Existence of ultraslow transverse modes is revealed and their
benefits in enhancement of pulse storage capacity is pointed out.
Finally, in our new results, we have introduced nonlinear optical
response and showed that it is possible to compensate dispersion
of a short ultraslow pulse for a specifically chosen peak power of
the pulse. This may be more efficient in optical storage than
group velocity dispersion by the Rabi frequency of the control
field. We hope these studies complement ongoing experimental
efforts for practical storage of coherent optical information in
atomic Bose-Einstein condensates.

\section{Acknowledgements}

O.E.M. gratefully acknowledges support from a T\"UBA/GEB\.{I}P
grant. We thank N. Postacioglu for many fruitful discussions and
for help in numerical computations.

%%%%%%%%%%%%%%%%%%%%%%%%%%%%%%%%%%%%%%%%%%%%%%%%%%%%%%%%%%%%%%%%%%%%
%%%%%%%%%%%%%%%%%%%%%%%%%%%%%%%%%%%%%%%%%%%%%%%%%%%%%%%%%%%%%%%%%%%%
%%%%%%%%%%%%%%%%%%%%%%%%%%%%%%%%%%%%%%%%%%%%%%%%%%%%%%%%%%%%%%%%%%%%

\end{document}